# Interfacing adaptive optics simulations with the optical model: a powerful tool for MORFEO

G. Pariani[*][a], G. Agapito[b], D. Magrin[c], M. Munari[d], L. Busoni[b], M. Riva[a], A. Di Rocco[e], P. Ciliegi[e]

[a]INAF – Osservatorio Astronomico di Brera, via E. Bianchi 46, 23807 Merate, Italy;
[b]INAF – Osservatorio Astronomico di Arcetri, Largo Enrico Fermi 5, 50125 Firenze, Italy.
[c]INAF – Osservatorio Astronomico di Padova, Vicolo dell'Osservatorio 5, 35122 Padova, Italy;
[d]INAF – Osservatorio Astrofisico di Catania, Via S. Sofia 78, 95123 Catania, Italy;
[e]INAF – Osservatorio di Astrofisica e Scienza dello spazio, Via Piero Gobetti 93/3, 40129 Bologna, Italy;

**ABSTRACT**

In the framework of the MORFEO project, the Multi-Conjugated Adaptive Optics (MCAO) module for the European Extremely Large Telescope (ELT), we developed an integrated modeling tool to interface the optical model with the adaptive optics simulations, called ASSO (Adaptive opticS Simulation interfaced with Optical model). This tool is our *asso nella manica* (ace in the hole) to predict the performances of the AO relay, i.e., to estimate the wavefront error within the technical and scientific fields of view after AO correction. The tool is based on the IDL based simulator PyrAmid Simulator Software for Adaptive opTics Arcetri (PASSATA), on Zemax OpticStudio for the optical modelling, and on Matlab as interface software.

## 1. INTRODUCTION

MORFEO[1], formerly known with the acronym MAORY, is the Multi-Conjugated Adaptive Optics (MCAO) module for the European Extremely Large Telescope (ELT). MORFEO is equipped with two post-focal deformable mirrors (DMs) conjugated at 6.5 and 17 km above the telescope and is designed to feed the Near Infrared (NIR) camera MICADO with both MCAO and Single-Conjugated AO (SCAO) operation modes.

In this paper, we present the integrated modeling tool we developed to interface the optical model with the adaptive optics simulations, called ASSO (Adaptive opticS Simulation interfaced with Optical model). This tool is our *asso nella manica* (ace in the hole) to predict the performances of the adaptive optics relay, i.e., to estimate the wavefront error within the technical and scientific fields of view after AO correction. The tool is based on the IDL PyrAmid Simulator Software for Adaptive opTics Arcetri (PASSATA)[2], on Zemax OpticStudio for the optical modelling, and on Matlab as interface software.

The general procedure is as follows: given the optical model, we calculated the wavefront maps for the laser guide stars (LGSs) and the natural guide stars (NGSs) with Zemax and Matlab; we computed the required DM corrections with the IDL based simulator PASSATA; we feed back the mirror surfaces in Zemax after the proper shape transformation; we evaluated the resulting wavefront map after AO correction.

After the setup of the interfaces and procedures between the different software, we performed a series of verifications to ensure the proper functioning of the machine. With this procedure we evaluated the relay performances given the optics manufacturing tolerances in terms of both low and high spatial frequency errors, considering the offload to the alignment optics and to the DMs. This last point is of crucial importance in the current phase of the design of MORFEO[3,4] to estimate part of the wavefront error budget and the level of absorption of any non-compliance of the relay by M4 or the MORFEO DMs.

*giorgio.pariani@inaf.it; phone +39 02 7232 0477

## 2. OPTICAL DESIGN AND WORKING PRINCIPLE

The main function of MORFEO is to relay the light beam from the ELT focal plane to the client instruments, while compensating the effects of the atmospheric turbulence and of the other disturbances of the wavefront. The present configuration passed the Preliminary Design Review (PDR) with ESO in 2021. In 2024 the consortium passed the Optical Critical Design Review (OCDR), granting the green light to issue the contracts for the final design and construction of the optomechanics.

MORFEO optical design is well described elsewhere[5,6] and is reported in Figure 1. A simple description is provided hereafter, to help the reader follow the discussion. MORFEO main path optics (MPO) provides an almost one-to-one image of the ELT focal surface to MICADO[7] entrance focal plane and to a second instrument port focal plane. The optical system accommodates two post focal Deformable Mirrors (DMs), conjugated at altitudes of 17.5 km and 6.5 km, that, together with the ELT M4, provide the multi-conjugated AO correction. Before the exit focal plane, the light of three Natural Guide Stars (NGSs) is picked up in the technical field of view to correct the low order WFEs. In the main optical path, near the pupil, a dichroic filter splits the scientific light from the Laser Guide Stars (LGSs) light, generated at the atmosphere sodium layer. The LGS light is then focused on the entrance focal plane of the LGS WaveFront Sensor (WFS) module by an objective.

The corrected Field of View of the instrument (FoV) is 160 arcsec, the central 80 arcsec being the scientific FoV and the external annular region the technical FoV. Each NGS arm is equipped with a 8×8 subaperture reference Shack-Hartmann Sensor (SHS), known as Reference WFS (RWFS), that can measure about a few tens of modes in a bandwidth of 600 – 1000 nm (more details can be found in the literature[8,9]). The main purpose of the three SHSs is a slow truth sensing to correct for any aberration in the science path up to the interface between MORFEO and MICADO and for any differential aberration between laser and natural light paths.

The LGS WFS module is equipped with focus adjustment and a derotator to compensate for sodium altitude variations and track telescope elevation. Each of the six LGS arm is equipped with a 68×68 subaperture SHS with a field of view of 16 arcsec, a framerate of 500Hz, and a pupil stabilization device, while the uplink jitter will be compensated by fast steering mirrors inside the launcher telescopes.

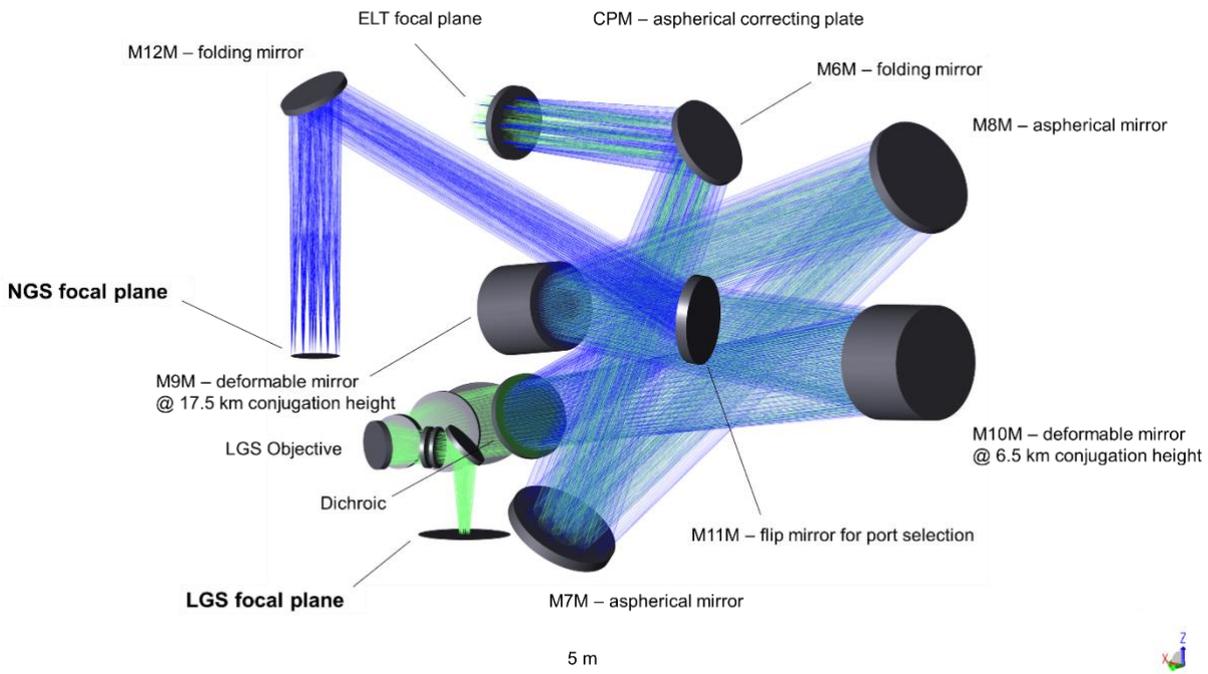

Figure 1: 3D view of the MORFEO optical configuration. Blue rays represent the science and NGS path to MICADO, cyan rays represent the LGS path.

The LGS asterism radius is kept fixed at 45 arcsec for any telescope elevation. The commands from the NGS truth sensing are obtained by a tomographic reconstruction[10] and are applied as bias to the high order (aka LGS) correction on the three DMs of MORFEO.

## 3. SOFTWARE ARCHITECTURE

The tool is based on Zemax OpticStudio for the optical modelling of the relay system. A given input file can be chosen, which may include manufacturing errors, thermoelastic movements, misalignments, etc. A Matlab based script is used to call Zemax and to return the performances of the relay before the correction (step 1). In a similar manner, the WFEs are calculated at the different NGS and LGS positions as given by the asterism and saved as .fits files (step 2). To this respect, an interface software developed in Matlab opens the desired Zemax file and calls the procedure to obtain the WFE maps. The maps are processed by the IDL based PyrAmid Simulator Software for Adaptive opTics Arcetri (PASSATA) and the corrections at the DMs conjugation height are calculated and stored as .fits files (step 3). The maps are then processed in Matlab to produce the mirror commands and are applied to the DMs in Zemax as grid sag surfaces (step 4). Again, the process is handled with Matlab and fully automatized. As for step 1, a Matlab based script returns the performances of the relay after the correction (step 5).

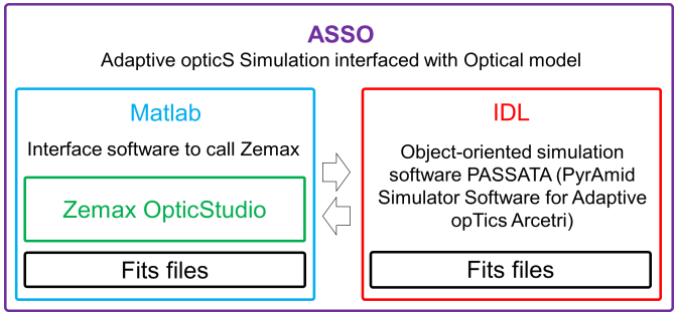

Figure 2: architecture of the ASSO software.

**STEP 1: calculation of the performances**

During the last few years, we developed a Matlab based environment to automatically analyze the performances of a given Zemax input file[5,6]. The tool is based on the ZOS-API programming interface, made available by OpticStudio. We use this interface to calculate all the relevant performance parameters of the relay system, from the WFE, to the distortion, to the footprint positions on the optics, and so on.

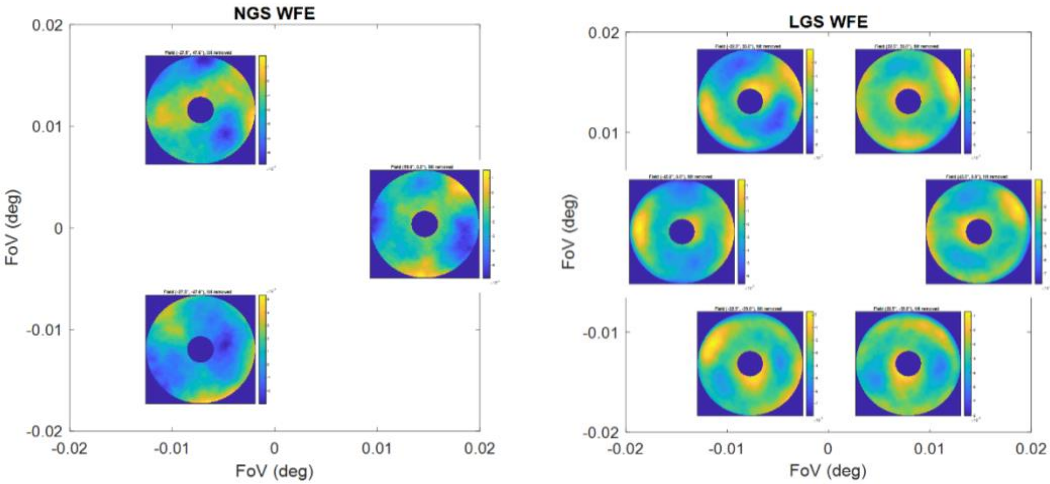

Figure 3: example of WFE maps at the NGS (left) and LGS (right) asterisms.

**STEP 2: calculation of the wavefront maps**

The WFE maps are calculated in Zemax for the three field positions of the NGS asterism at the exit focal plane, and for the six field positions of the LGS asterism at the exit focal plane of the LGS objective (Figure 3). In both cases, we used the rectangular array grid, 32x32, and we referred the WFE to the exit pupil.

Concerning the tilt component of the WFE, we usually refer the WFE to the centroid. This is formally correct for the LGS wavefront sensor (WFS), used to correct for high orders only; for the NGS WFS, considering the WFE without tilt is equivalent to optimize the WFE for all the low order errors except the plate scale.

We could I principle refer the WFE to the chief ray to directly account for WFE tilt in Zemax, but we observe problems when a grid sag is applied on the mirrors. This is because the rays are spread around, and usually the chief ray is far apart and does not represent a meaningful reference. We accordingly calculate the WFE as the sum of the WFE referenced to the centroid, e.g., tip/tilt removed, and the tilt contribution given by the shift in the centroid between the actual configuration and a given reference configuration (the nominal, e.g.). The tilt coefficients are according to eq. 1 and 2:

$$tx = \frac{1}{2}(x - x_{ref})\frac{EPD_{ref}}{EFL_{ref}} \qquad (1)$$

$$ty = \frac{1}{2}(y - y_{ref})\frac{EPD_{ref}}{EFL_{ref}} \qquad (2)$$

over a pupil of radius 1. We consider as effective focal length (EFL) and entrance pupil diameter (EPD) the ones of the reference configuration (Figure 4 shows an example).

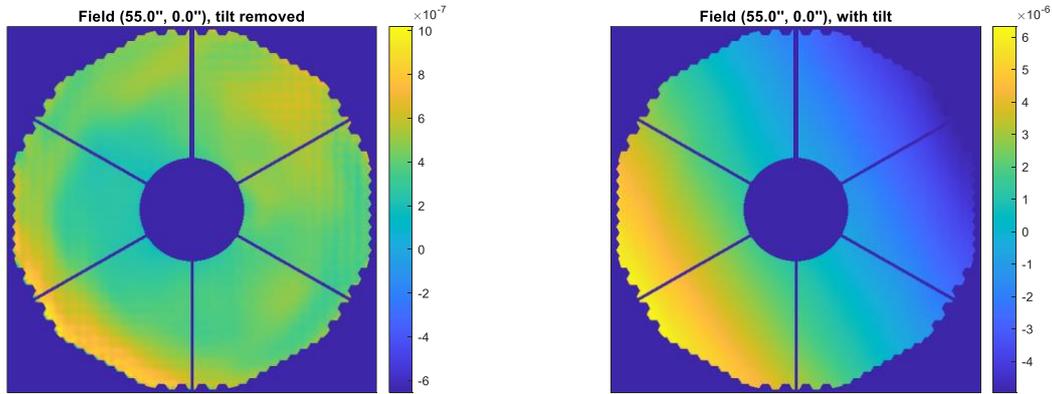

Figure 4: example of a NGS pupil map referenced to the centroid, i.e., without tilt (left), and with tilt (right)

**STEP 3: calculation of correction maps**

Starting from the phase maps a short PASSATA simulation is performed. The simulation is short because there is no atmospheric turbulence and the LGS WFSs and RWFSs only sense the aberration coming from the optical model, so the convergence to the steady state is fast (note that in presence of atmospheric turbulence the truth sensing is very slow with a temporal bandwidth of tenths of Hz[10]). The DM commands are stored to retrieve the corresponding phase map $\varphi(\vartheta,\xi)$ that are shared as .fits files.

**STEP 4: application of mirror commands**

With a Matlab script we transform each phase map $\varphi(\vartheta,\xi)$ calculated on sky in the mirror command, to be applied on the optics surface as sag $z(x,y)$, according to eq. 3:

$$z(x,y) = -\frac{M \cdot \varphi(\vartheta,\xi)}{2\cos[\alpha(\vartheta,\xi)]} \qquad (3)$$

where $M$ is the rototranslation matrix between the coordinates on sky and the coordinates on the mirror and $\alpha$ is the local incidence angle of the chief ray. Since Zemax requires a regular grid for the sag application, we practically proceed as follows:
- we calculate for any position on sky $(\vartheta,\xi)$ the corresponding position $(x,y)$ on the conjugated optics and the incidence angle of the chief ray;
- we set a regular grid of coordinates on the optics;
- we calculate the corresponding positions on sky;
- we fit the on-sky map on the new coordinates;

If a grid sag map is present on the optics, we automatically read the sag on the same regular grid and we apply the sum of the two.

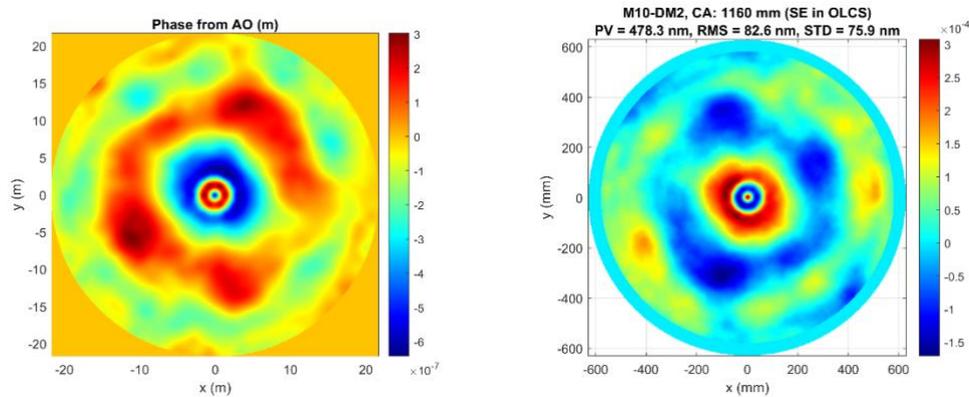

Figure 5: example of a phase map on sky (left) and of the corresponding mirror command on DM2 (right).

**STEP 5: calculation of the performances**

As for step 1, all the performances can be evaluated again on the modified file.

## 4. SOFTWARE VERIFICATIONS

Once set the interface between the difference software, the main verifications consisted in:
- phase map orientation and scaling;
- command application (in presence and absence of WFE tilt).

Concerning the first point, we followed two parallel procedures. In the former procedure, we constructed a Zemax file with low order aberrations on the each deformable mirror (M4, M9M, and M10M). We followed steps 2-4 of the procedure and we evaluated the residuals, as shown in Figure 6. Any major error in the map rototraslation and invertion has been highlighted and corrected.

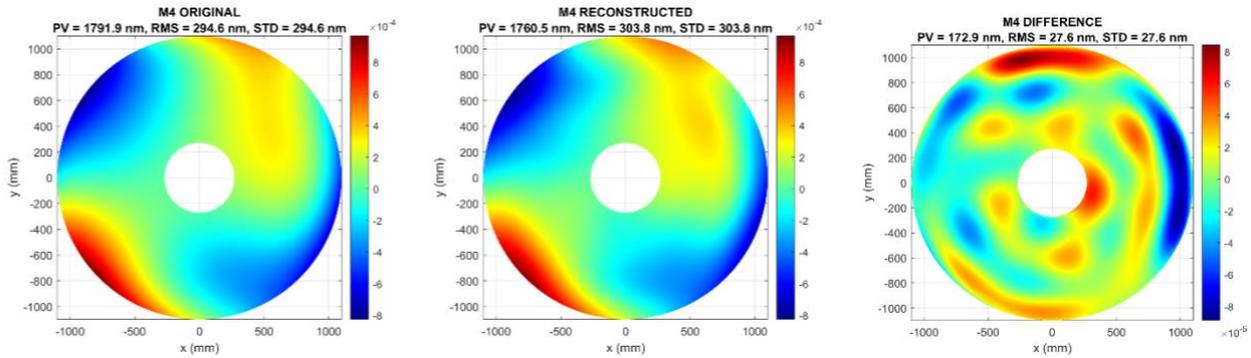

Figure 6: application of low order aberrations on M4: original mirror sag (left), reconstructed sag (center), and sag difference (right).

In the latter procedure, we started from a WFE with local aberrations on sky (simulating the poke of three DM actuators). We processed the phase map as for step 4 and we calculated the WFE at the NGS asterism locations. Then, we compared the positions of the peaks in the WFE, finding a good match or discrepancies within one pixel level (Figure 7).

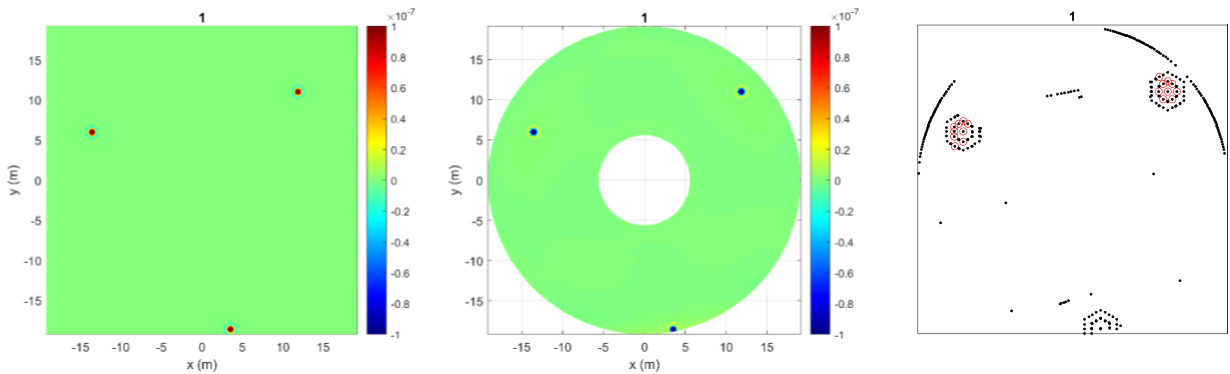

Figure 7: application of local aberrations on M4: original WFE on sky (left), calculated WFE (center), and peak identification – red circles mark the positions where the peaks in the two WFE maps correspond (right).

Concerning the verification of the scaling when applying a command, we proceeded as follows for each low order Zernike mode:
- we created a Zemax file with a Zernike standard sag surface on the desired DM, and we applied a single mode;
- we followed steps 2-4 to obtain the corresponding mirror command, WFE is calculated with tilt;
- we compared the applied surface sag with the mirror command: a perfect match would result in a null map.

Since the procedure aims at minimizing the WFE on the positions of the asterisms' stars, the calculated mirror command includes also higher order terms. In this respect, we need to compare the mirror command in the different cases with the mirror command in the nominal case to get rid of this effect. As shown in Figure 8, the mirror command difference resembles very well the applied mirror sag up to the astigmatism, and any difference has been verified to be the effect of the gain we apply in the AO simulator.
The coma, instead, has a strong coupling with the tilt: even if applied as pure coma, it is seen by the simulator as a coma plus tilt, since the centroid is shifted from the chief ray. Accordingly, the actual command is a mix of tilt and coma.

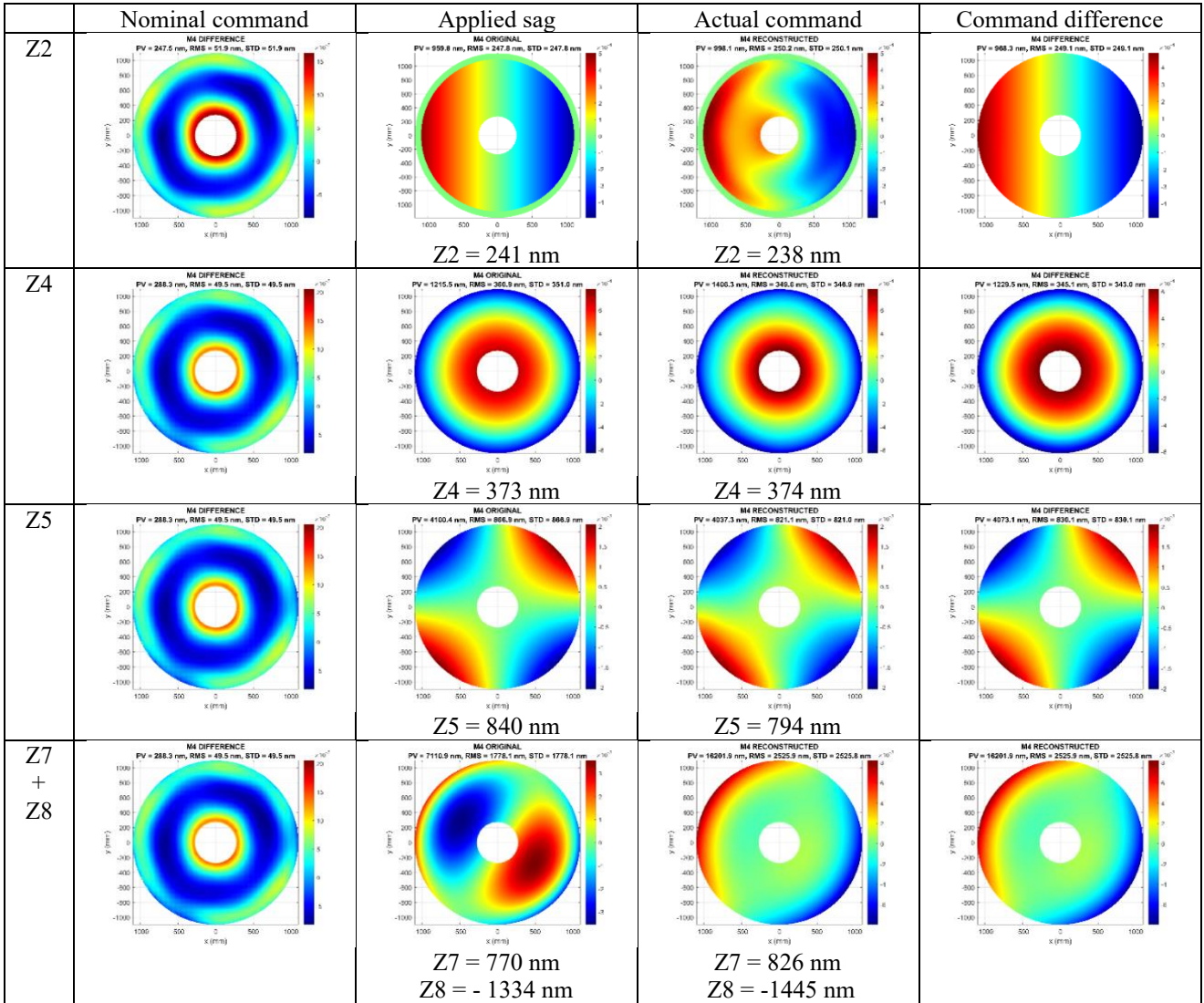

Figure 8: application of low order Zernike terms on M4. The command difference (last column) is the difference between the actual command and the nominal command.

To better understand this effect, we performed an iterative correction, as reported in Figure 9. We considered the nominal design as reference for the centroid calculation. The residual tilt converges to zero in a dozen iterations. Accordingly, the mirror command converges to a pure coma, and the residuals to zero. Worth noting that the WFE converges to a stable value in just one iteration, meaning that the residuals in the mirror command are mainly tilt; in fact, the corresponding residuals in the centroid tilt is not field dependent.

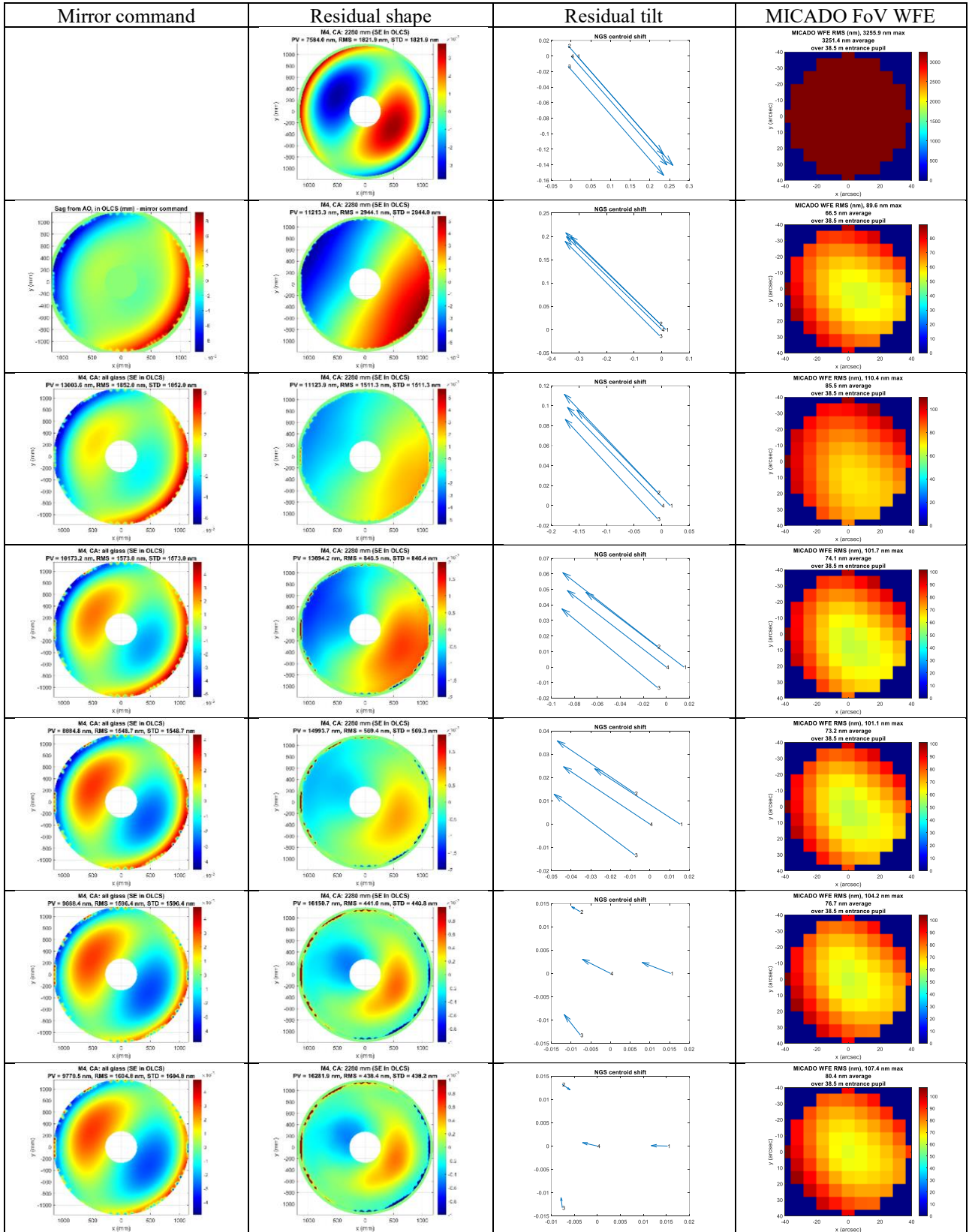

Figure 9: iterative application of coma on M4.

# 5. A STUDY CASE

Given the freedom on the Zemax file used as input, a series of optical effects can be simulated with ASSO. We considered all the possible cases related to the verification of the instrument requirements, which will be requested by FDR. In particular, the following:

- optics manufacturing tolerances, in terms of both low and high spatial frequency errors;
- optics and mechanics thermoelastic effects, for both the optical mountings and the mechanical structure;
- optics misalignments for the initial instrument alignment and for stability considerations.

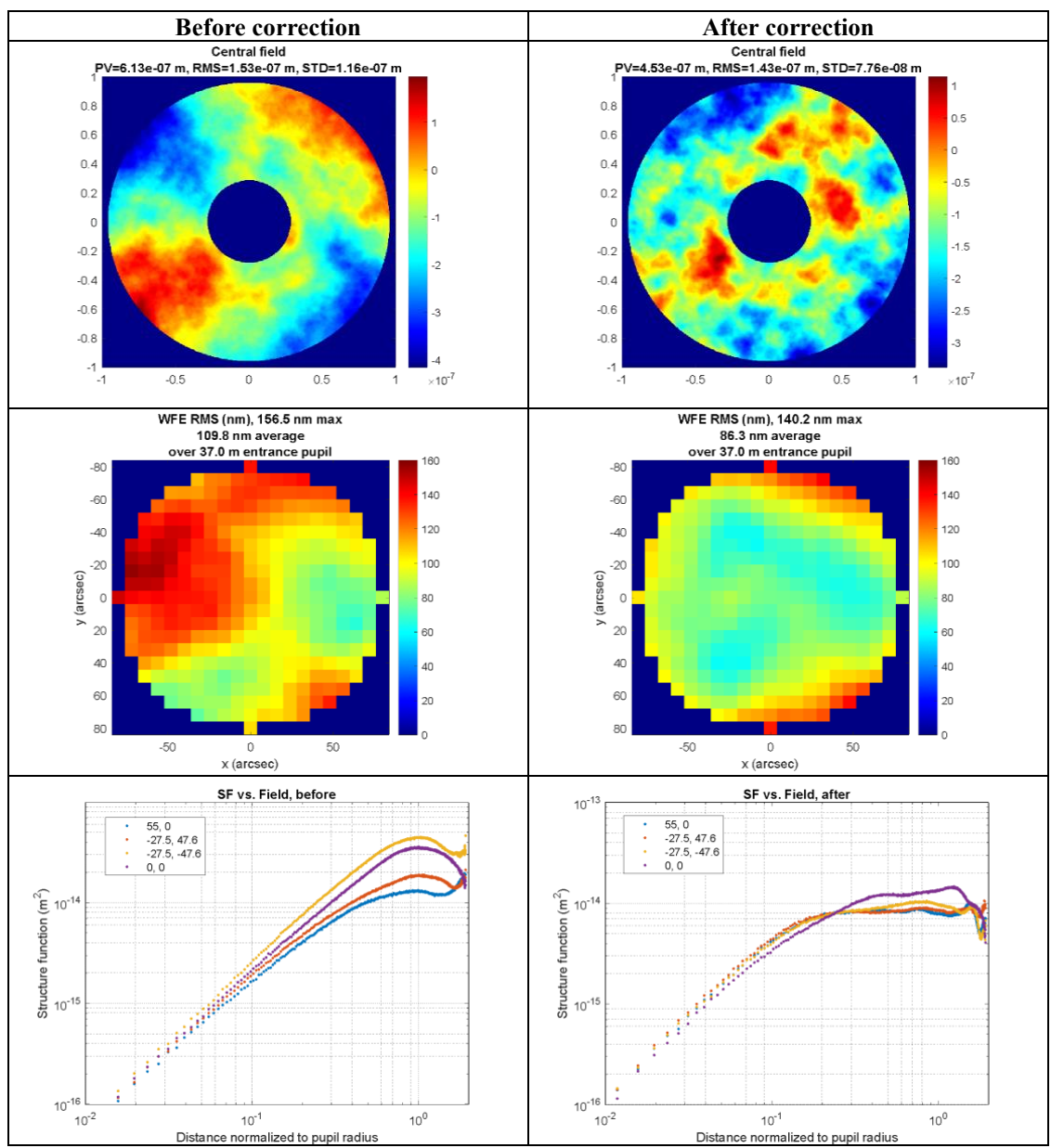

Figure 10: On-axis WFE in the MORFEO exit pupil (top), WFE RMS in the technical FoV (middle), and structure functions on-axis and in the three NGSs positions (bottom), before (left) and after (right) the processing with ASSO.

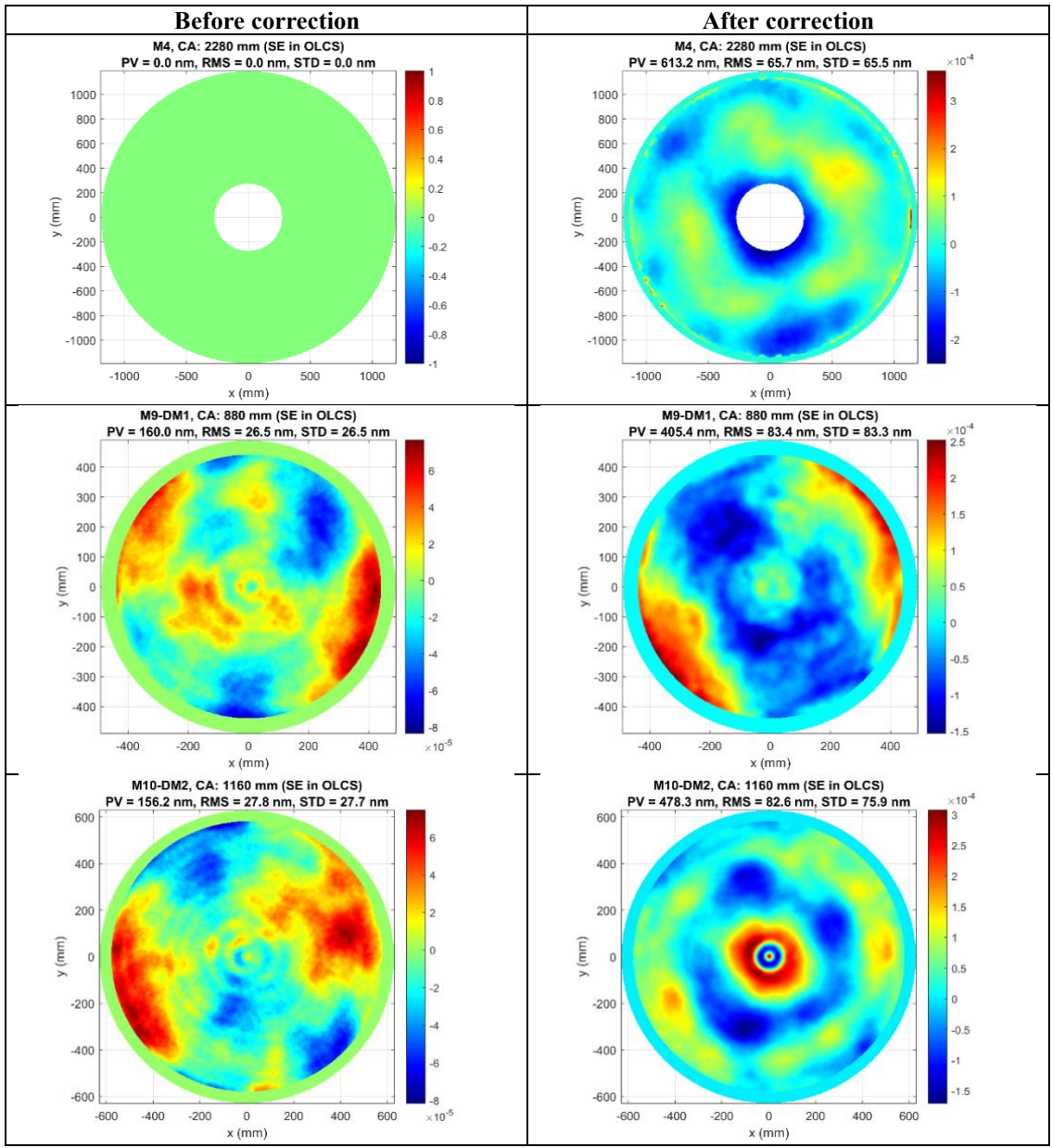

Figure 11: surface sag of the three deformable mirrors before (left) and after (right) the processing with ASSO.

As an example, we present the case of the manufacturing tolerances of the optics, in terms of mid/high spatial frequency errors. We described each surface by Zernike polynomials with random coefficients up to Z=36 (in the Noll notation), and by a surface generated from a given PSD for the higher order errors. Details on the procedure can be found in the literature[6]. Accordingly, we described each optical surface in Zemax as a grid sag surface, with a frequency corresponding to twice the maximum frequency of the PSD (400 m$^{-1}$ = 2.5 mm); considering the size of MORFEO elements, we end up with maps of the order of 500x500 pixels. We applied these surfaces to each optics and we followed the procedure described above. Figure 10 reports the on-axis WFE, the WFE RMS in the technical FoV, and the structure function of the WFE before and after the correction. The astigmatism present in the on-axis WFE is corrected by the DMs, leaving only higher order errors. Similarly, the RMS WFE improves especially in the central region, being the scientific FoV. It moves from 110 to 86 nm

RMS in the technical FoV, and from 110 to 74 nm RMS in the scientific FoV. Worth noting that the best correction is in the position of the three NGSs and not on-axis. This is evident in the WFE RMS map and in the structure functions. The structure functions show also that the AO correction is very effective for spatial frequencies above 1/10 of the pupil radius. The offload on the three deformable mirrors, M4, DM1, and DM2, is reported in Figure 11. In general, we observed maximum offloads of few hundreds of nm RMS, which is negligible compared to the admissible stroke of the DMs.

## 6. CONCLUSIONS

We described the integrated modeling tool, called ASSO (Adaptive opticS Simulation interfaced with Optical model), we developed to interface the optical model with the adaptive optics simulations for the MORFEO instrument. The tool is based on the IDL based adaptive optics simulator PyrAmid Simulator Software for Adaptive opTics Arcetri (PASSATA), on Zemax OpticStudio for the optical modelling, and on Matlab as interface software.

We showed the verifications we performed to guarantee proper communication between the different software, in particular in terms of map exchange (WFEs and phase screens) and command application process (in presence and absence of WFE tilt).

As an example, we reported the results we obtained applying mid/high spatial frequency surface errors to each optics, and we showed the WFE performances of the relay before and after the AO correction. We ensured that the required DM correction is well within their admissible stroke. This example is representative of the possible scenarios that will take place during the optics procurement, when the impact of any non-compliance shall be evaluated to possibly relax the optics specifications or to estimate the level of absorption of the non-compliance on M4 and on the MORFEO DMs.

This work is also part of the wavefront error budget estimate that is used to predict the overall performance of MORFEO, presented in another paper at this conference[4].